\documentclass[reprint,aps, prm,superscriptaddress,amsmath,amssymb]{revtex4-2}
\usepackage{graphicx}
\usepackage{dcolumn}
\usepackage{bm}
\usepackage[mathlines]{lineno}
\usepackage{etoolbox}
\usepackage{gensymb}
\usepackage{graphicx}
\usepackage{subfigure}
\usepackage{amsmath}
\usepackage{amssymb}
\usepackage{dsfont}
\usepackage{hyperref}
\usepackage{multirow}
\usepackage{rotating}
\usepackage{epstopdf}
\usepackage{color}
\usepackage{wrapfig}
\usepackage{fancyhdr}
\usepackage{lipsum}
\usepackage{tabularx}
\usepackage[capitalize]{cleveref}

\usepackage[T1]{fontenc}
\DeclareUnicodeCharacter{2212}{-}
\DeclareUnicodeCharacter{03B4}{$\delta$}

\usepackage[labelfont=bf,labelsep=quad,justification=justified]{caption}

\begin{document}	
	\preprint{APS}	
	\title{Revisiting the magnetic ordering through anisotropic magnetic entropy change in
		quasi-two-dimensional metallic ferromagnet, Fe$_4$GeTe$_2$}
	\author{Satyabrata Bera}
	\affiliation{
		School of Physical Sciences, Indian Association for the Cultivation Of Science, Jadavpur, Kolkata-700032
	}
	
	\author{Suman Kalyan Pradhan}
	\affiliation{
		School of Physical Sciences, Indian Association for the Cultivation Of Science, Jadavpur, Kolkata-700032
	}
	\author{Md Salman Khan}
	\affiliation{
		School of Materials Sciences, Indian Association for the Cultivation Of Science, Jadavpur, Kolkata-700032
	}
	\author{Riju Pal}
	\affiliation{
		SN Bose National Centre for Basic sciences, JD BLOCK SECTOR III, SALT LAKE (NEAR WATER TANK 14),Kolkata - 700106
	}
	\author{Buddhadeb Pal}
\affiliation{
	SN Bose National Centre for Basic sciences, JD BLOCK SECTOR III, SALT LAKE (NEAR WATER TANK 14),Kolkata - 700106
}
	\author{Sk Kalimuddin}
	\affiliation{
		School of Physical Sciences, Indian Association for the Cultivation Of Science, Jadavpur, Kolkata-700032
	}
	\author{Arnab Bera}
	\affiliation{
		School of Physical Sciences, Indian Association for the Cultivation Of Science, Jadavpur, Kolkata-700032
	}
	\author{Biswajit Das}
	\affiliation{
		School of Physical Sciences, Indian Association for the Cultivation Of Science, Jadavpur, Kolkata-700032
	}
	\author{Atindra Nath Pal}
	\affiliation{
		SN Bose National Centre for Basic sciences, JD BLOCK SECTOR III, SALT LAKE (NEAR WATER TANK 14),Kolkata - 700106
	}
	\author{Mintu Mondal}%
	\email{sspmm4@iacs.res.in}
	\affiliation{
		School of Physical Sciences, Indian Association for the Cultivation Of Science, Jadavpur, Kolkata-700032 
	}%
	
	\date{\today}

\begin{abstract}
	We have investigated the nature of ferromagnetic order and phase transitions in two dimensional (2D) van der Waals (vdW) layered material, Fe$_4$GeTe$_2$ through measurements of magnetization, magneto-caloric Effect (MCE), and heat capacity. Fe$_4$GeTe$_2$ hosts a complex magnetic phase with two distinct transitions: paramagnetic to ferromagnetic at around $T_\text{C}$ $\sim$ 266 K and another spin reorientation transition (SRT) at around $T_\text{SRT}$ $\sim $ 100 K. The magnetization measurements shows a prominent thermal hysteresis in proximity to $T_\text{SRT}$ at $H\parallel c$, which implies the first-order nature of SRT. 
	 Reasonable MCE has been observed around both transition temperatures ( at around $T_\text{C}$, -$\Delta$S$_M^\text{max}$ = 1.95  and 1.99 J.Kg$^{-1}$K$^{-1}$ and at around  $T_\text{SRT}$, -$\Delta$S$_M^\text{max}$= 3.9 and 2.4 J.Kg$^{-1}$K$^{-1}$ along $H\parallel ab$ and $H\parallel c$  respectively) at 50 kOe magnetic field change. The above results reveal higher MCE value at $T_\text{SRT}$ compared to the values of  MCE at $T_\text{C}$. The scaling analysis of MCE at $T_\text{C}$, shows that the rescaled $\Delta$S$_M (T, H)$ follow a universal curve confirming the second-order character of the ferromagnetic transition. The same scaling analysis of MCE breaks down at $T_\text{SRT}$ suggesting that SRT is not a second order phase transition. The exponent $n$ from field dependence of magnetic entropy change presents a maximum of  $|n|>2$ confirming the first-order nature of SRT.
\end{abstract}
\maketitle
{\section{Introduction}}

The intrinsic long-range ferromagnetic order in two-dimensional (2D) van der Waals (vdW) materials has recently generated immense interest for both studying the fundamental science of magnetism in low-dimensional systems and technological applications for spintronics \cite{Ding2021, Huang2017, Gong2017, Burch2018}. Moreover,  these layered materials allow them to be cleaved down to monolayers and amalgamate with other 2D materials to create novel heterostructures without any lattice similarity and provide versatile scientific platforms for studying various novel quantum phenomena\cite{Geim2013, Geim2007, Novoselov2005}. However,  the typical Curie temperatures of these magnetic materials are low compared to their three-dimensional (3D) counterparts \cite{Kabiraj2020}. Therefore, finding new layered 2D ferromagnetic materials having higher Curie temperatures led to new directions and the discovery of many new materials\cite{Jiang2018,Zhang2019,McGuire2017}. The recent addition to this family of 2D magnetic materials is Fe-based Germanium Telluride layered compounds (FGT), Fe$_l$GeTe$_2$ ($l$ = 3, 4, and 5) \cite{Kim2018,Seo2020,May2019}. The flexibility of tuning the chemical composition results in varieties of complex magnetic orders and physical properties based on the value of $l$ in Fe$_l$GeTe$_2$ which is emerging as model systems for studying low dimensional magnetic ordering and phase transitions. 

Recent studies on Fe$_4$GeTe$_2$ (F4GT) compound reveal reasonably large $T_\text{C}$  along with a complex magnetic state \cite{,Seo2020,Mondal2021}.The theoretical studies indicate that the perpendicular electric field could enhance the in-plane and out-of-plane magnetic anisotropies in F4GT mono-layers with significant potential practical applications \cite{Kim2021}. In addition, the first-principles calculations and the group theory analysis have predicted large anomalous Nernst and thermal Hall conductivities in bi-layer ferromagnetic F4GT \cite{Yang2021}. The near room temperature ferromagnetism with the significant magnetic moment and large conductivity make F4GT a promising 2D magnetic material for magneto-optical devices, spintronics, and spin caloritronics applications\cite{Seo2020, Mondal2021}.

The Merin-Wagner theorem says that no long-range magnetic ordering is possible in the 2D limit in isotropic magnetic materials \cite{Mermin1966}. However, the anisotropy can open a gap in magnon spectra leading to the magnetic ordering in these compounds \cite{Li2020}. In general, the anisotropy in a magnetic material is responsible for the various intrinsic magnetic properties, which are mainly of three types: (i) magnetocrystalline anisotropy, (ii) shape anisotropy, and (iii) exchange anisotropy. The magnetocrystalline anisotropy play significant role in magneto-caloric effect (MCE)\cite{Fries2016,Caron2013}. Therefore, investigation of magnetization combined with measurement of MCE can reveal the nature of magnetic ground state and phase transitions\cite{Liu2018b,Liu2019,Liu2020a}. In addition, the MCE has also been used as a promising environmentally-friendly cooling solution different from conventional gas compression cooling and attracted immense research interest over the past few decades \cite{Franco2018,GschneidnerJr2005}. The critical behavior of MCE in the proximity of second-order magnetic phase transition is independent of the microscopic details and determined by the dimensionality, range of interaction, and symmetry of the order parameter. The critical exponent analysis based on two concepts: universality\cite{Griffiths1970} and scaling theory \cite{Kadanoff2000}, can reveal the nature of phase transitions. Unambiguous determination of the order of ferromagnetic and spin reorientation transitions are essential for both sciences of magnetism in low dimensional systems and technological applications\cite{Law2018}. However, the measurement of MCE in F4GT compound and the validation of universality in terms of critical scaling analysis of MCE data have not yet been reported.

Here we report a detailed investigation of magnetism and magneto caloric effect (MCE) in F4GT, revealing an interesting magnetic phase with two distinct transitions: paramagnetic to ferromagnetic at $T_\text{C}$ $\sim$ 266 K and another spin reorientation transition at around $T_\text{SRT}$ $\sim $ 100 K. The temperature-dependent magnetization shows a prominent thermal hysteresis at around $T_\text{SRT}$ indicating the first-order nature of the transition.The MCE has been estimated from direction-dependent magnetization measurements at around $T_\text{SRT}$ and $T_\text{C}$. Obtained MCE values are moderate in terms of application point of view. The MCE value at around  $T_\text{C}$ is almost isotropic, while it shows strong anisotropy at around $T_\text{SRT}$, reconfirming the first-order nature of the transition at around $T_\text{SRT}$. We have also done the scaling analysis of magnetic entropy change (-$\Delta$S$_M$) in proximity to ferromagnetic transition at $T_\text{C}$ and estimated the critical exponents to describe the nature of magnetic exchange interaction and symmetry of the order parameter. The universal scaling of MCE  is independent of temperature ($T$) and magnetic field ($H$) and confirms the second-order nature of the ferromagnetic transition. Whereas the scaling analysis of  -$\Delta$S$_M$ in proximity to SRT, confirms the first-order nature of the transition. In this study, we have unambiguously confirmed the second-order nature of ferromagnetic transition at $T_\text{C}$ and the first-order nature of spin reorientation phase transition at $T_\text{SRT}$.\\

\begin{figure}
		\centering
		\includegraphics[width=1\columnwidth]{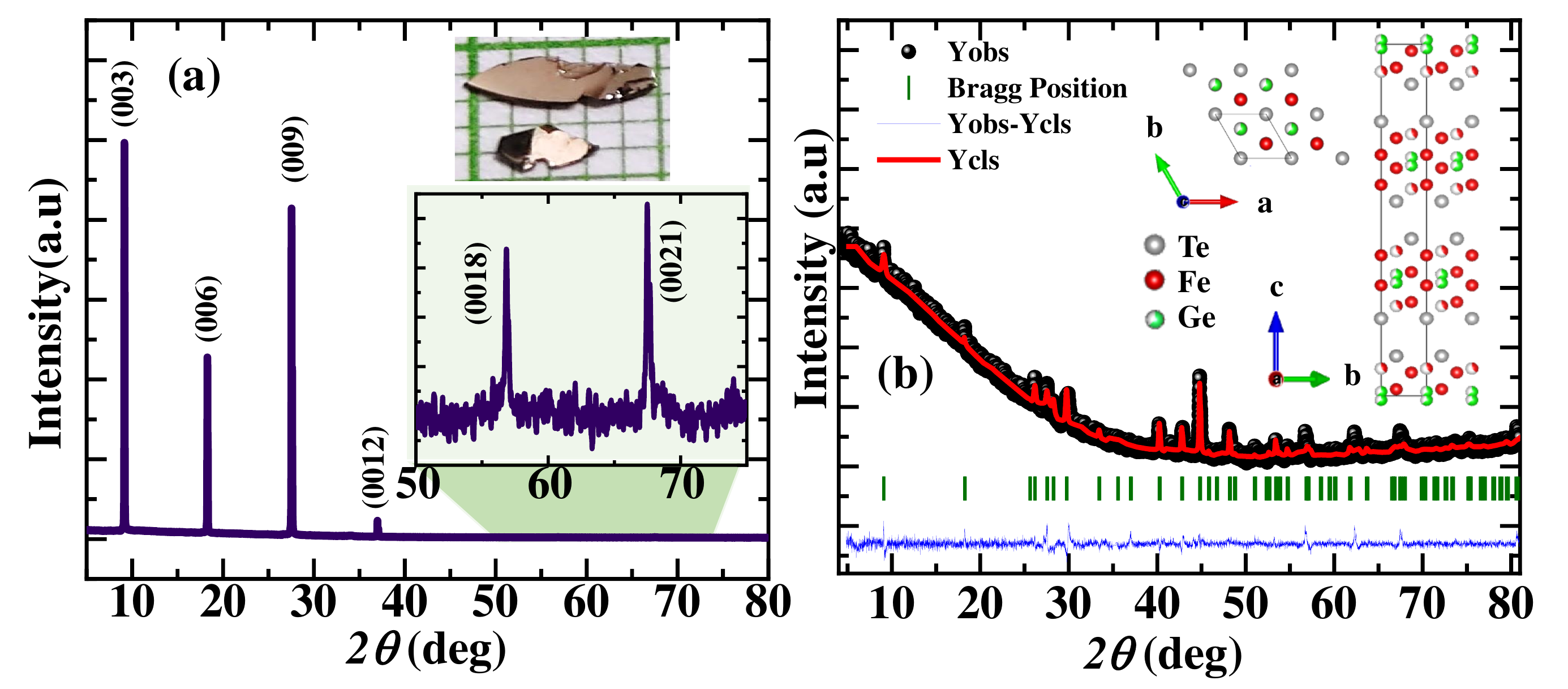}   	
		\caption{\textbf{Structural analysis of F4GT (F4GT)crystal.} (a) The X-ray diffraction pattern obtained from the cleaved plane of F4GT single crystal at room temperature. The picture of the single crystal is shown in the inset. (b) Observed (solid black sphere) and calculated (solid red line) powder X-ray-diffraction patterns for the ground F4GT single crystals. The small green lines denote the Bragg positions, and the blue curve indicates the difference between the observed and calculated patterns. View of the crystallographic structure of F4GT from the a axis and c axis. The ash,  red,and green spheres represent the Te, Fe and Ge respectively. The black box with the cross-section rectangles depicts a crystallographic unit cell.}
		\label{fig:XRD}	
	\end{figure}

	{\section{EXPERIMENTAL DETAILS}}
High-quality single crystals of F4GT were grown by the chemical vapor transport (CVT) method using I$_2$ as a transport agent. The mixture of [Fe (99.99$\%$ pure), Ge (99.99$\%$ pure), and Te (99.99$\%$ pure)]  powders in a molar ratio Fe:Ge:Te = 4.5:1:2 along with the transport agent, was vacuum-sealed in a quartz tube. Excess Fe was used to maintain the stoichiometric ratio, as there is a natural tendency for having vacancy at the Fe site in FGT compounds. Then the vacuum-sealed quartz tube was placed inside a two-zone horizontal tube furnace with source and growth temperatures $\simeq$ 750$^\circ~C$ and 700$^\circ~C$, respectively. The temperatures were kept fixed with minimal temperature fluctuations ($\pm 1^\circ~C$), maintaining a constant temperature gradient for seven days to complete the growth process. After the growth process, the furnace temperature was reduced to room temperature (RT). After that, the quartz tube was broken under ambient conditions, and a few large, thin, shine-surface, good quality single crystals with lateral dimensions up to eight millimeters were obtained (see inset of Figure~\textcolor{blue}{\ref{fig:XRD}.(a)}). Preliminary X-ray diffraction (XRD) study of a small piece of the single crystal at room temperature  (RT) confirms the formation of F4GT samples. To determine the crystal structures of single-crystalline samples, powder XRD has been carried out using fine powder prepared by crushing a small piece of as-grown single crystal samples. The X-ray diffraction (XRD) data were taken with Cu K$_{\alpha}$ ($\lambda$= 1.54\AA) radiation by a Rigaku powder diffractometer. The weight percentage of the obtained phase was estimated by Rietveld fitting using FullProf Suite \cite{RodriguezCarvajal1993} (see Figure~\textcolor{blue}{\ref{fig:XRD}.(b)}). 

Crystal Morphology and chemical compositions were confirmed using a high-vacuum scanning electron microscope (SEM) (JEOL LSM-6500). Samples were placed on carbon tape before loading into the SEM chamber and examined under an accelerated voltage of 10 kV. Results are discussed in appendix \textcolor{blue}{\ref{apsec:SEM-EDX}} . The dc magnetization measurements were done in a Quantum Design Magnetic Properties Measurement System (MPMS-SQUID) with the applied magnetic field ($H$) up to 50 kOe. A piece of F4GT single crystal was mounted in the plastic pipe with $H$ parallel to the $ab$ plane and $c$ axis to measure the magnetic moment ($M$) as a function of $T$ and $H$. Heat capacity at zero and 20 kOe magnetic fields was measured using the Quantum Design Physical Properties Measurement System (PPMS).\\

{\section{Results \& Discussion}}
	\subsection{XRD Analysis and Crystal structure}
	
Figure.\textcolor{blue} {\ref{fig:XRD}(a)} presents the XRD pattern of F4GT single crystal at RT($T \simeq$ 298 K), all peaks come from (00\textit{l}), which indicates that the crystal surface is normal to the \textit{c} axis with the plate-shaped surface parallel to the \textit{ab} plane. A single crystal with moderate dimension of $\sim$ 5 $\times$ 2 $\times$ 0.3 mm$^{3}$  is shown in the inset of Figure.\textcolor{blue} {\ref{fig:XRD}(a)}. An excellent fit of the powder XRD pattern shown in Figure. \textcolor{blue}{\ref{fig:XRD}(b)}, indicates that F4GT crystallizes in a trigonal crystal symmetry with the space group $R$$\overline{3}$m (Space Group no. 166). The obtained lattice, refinement parameters, and atomic positions are presented in Table I and are well matched with the previous results \cite{Seo2020}. Inset of Figure.\textcolor{blue}{\ref{fig:XRD}(b)}  shows a projection  of  the  crystal  structure  of F4GT, emphasizing Fe and Ge atoms are sandwitched between layers of Te atom.
However, our system has two types of Fe atom (Fe2 and Fe3) with full occupancy, another kind of Fe atom (Fe1) exists, and one Fe atom (Fe1) is shared with the Ge atom. This is similar to what was reported in Fe$_5$GeTe$_2$(F5GT). But for this system, the obtained sharing ratio between Ge and Fe(1) is 0.8:0.2. This is way smaller than the F5GT phase, as reported in earlier literature\cite{May2019a}. The EDX results show stoichiometry of the sample is Fe$_{4.18}$Ge$_{0.95}$Te$_{1.92}$, which confirms the above observation (for details see appendix \textcolor{blue}{\ref{apsec:SEM-EDX}}).

	\begin{table}
		\caption{The Rietveld refined crystallographic parameters, obtained from X-ray diffraction study at $T$ = 300 K} 
		\begin{tabular}{|p{4cm}|p{4cm}|}
			\hline
			Nominal composition & F4GT  \\
			Refined composition & Fe$_{4.2}$Ge$_{0.8}$Te$_2$  \\
			Structure &  Rhombohedral \\
			Space group & $R$-3m (No. 166) \\
			Lattice parameters &  \\
			a(\AA) & 4.04411 \\
			b(\AA) & 4.04411 \\
			c(\AA) & 29.1402 \\
			V$_{cell}$(\AA$^3$) &412.73285 \\
		   $\chi^2$ & 1.7 \\
			\hline
		\end{tabular}\\
	\centering 
	\begin{tabular}{|p{1.05cm}|p{1.1cm}|p{1.1cm}|p{1.1cm}|p{1.5cm}|p{1.55cm}|}
			\hline
			Atom & Wyckoff position &$x$ & $y$ & $z$  & Occupancy \\
			\hline
			Te & 6c & 0 & 0 & 0.21986 & 1.0 \\
			Ge & 6c & 0 & 0 & 0.01179 & 0.725 \\
			Fe1 & 6c & 0 & 0 &  0.07791 & 0.275\\
			Fe2 & 6c & 0 & 0 & 0.31443& 1.0 \\
			Fe3 & 6c & 0 & 0 & 0.39972 & 1.0 \\
			\hline
		\end{tabular} 
		\label{table:XRD} 
	\end{table} 
\subsection{Magnetization}

\begin{figure}
	\centering
	\includegraphics[width=1\linewidth]{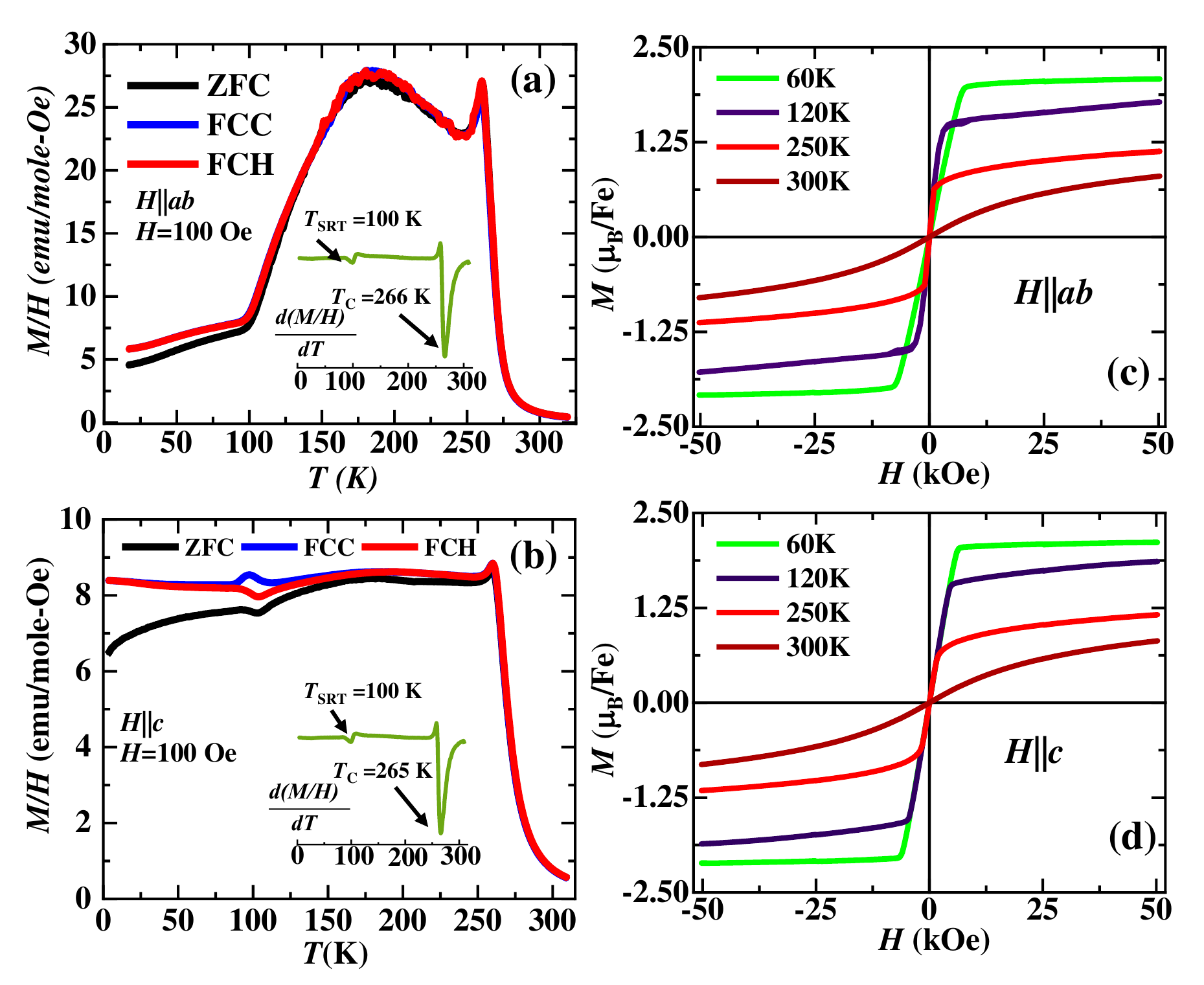}   	
	\caption{\textbf{Magnetization measurements.} (a-b)Temperature dependence of the dc magnetization ($M/H$), of single-crystalline F4GT at $\mu_0H$ = 0.1 kOe, 	measured in zero-field-cooled (ZFC-black solid line), field-cooled (FCC-solid blue line) and  field-cooled-heating (FCH-solid red line) protocols, for directions along $(H\parallel ab)$ and $(H\parallel c)$, respectively. Inset shows the 1$^{st}$ derivative of $(M/H)$. (c-d) Isothermal magnetizations: parallel and  perpendicular to the $ab$ plan, measured at $T$ = 60, 120, 250, and 300 K, respectively, up to $\mu_0H$ = 50 kOe.}
	\label{fig:MT}	
\end{figure} 

The magnetization measurement reveals a complex magnetic phase of F4GT with two distinct transitions. Zero-field-cooled (ZFC), field-cooled cooling (FCC), and field-cooled heating (FCH) magnetization ($M$) data of the as-studied sample were measured as a function of temperature ($T$) under an applied dc magnetic field ($H$) of 0.1 kOe in parallel to the $ab$ and $c$ directions and corresponding data are shown in Figure. \textcolor{blue}{\ref{fig:MT}(a-b)}. There are following few important observations as described below.\\
(1) The sharp increase of $M(T)$ with decreasing temperature which clearly suggesting a FM phase transition of the sample at $T_\text{C}$ $\sim$ 266 K, which matches well with the earlier literetures\cite{Seo2020,Mondal2021}[obtained from the $\frac{dM}{dT}|_{FCC}$ vs $T$ plot]. To estimate the effective magnetic moment of Fe atom, the $M/H$ vs $T$  curve of powder sample of F4GT has been analyzed using the  modified Curie-Weiss (CW) law and the estimated values of effective magnetic moment, $P_\text{eff}$ = 5.52 $\mu_B$/Fe with Curie-Weiss temperature,~$\Theta_p$~=~285.2 K (see the appendix~\ref{app:Magnetization}). The +ve value of $\Theta_p$ confirms the FM characteristics of F4GT.\\
(2) The ZFC curve shows a well-defined sharp peak corresponding to $T_\text{C}$ or just below it, similar to FCC and FCW. This is a typical example of Hopkinson maxima\cite{Slama2017}. Often soft magnet,such as ferrites, shows this kind of feature\cite{Saghayezhian2019}. \\
(3) Below $T_\text{C}$, the magnetic moment along the $ab$ plane decreases rapidly with decreasing temperature and becomes comparable to the magnetic moment along the $c$ axis. This is due to the fact that all \textit{Fe} moments are not aligned along the same direction, which implies some spins are canted in that temperature region\cite{May2019}. In addition, "cusp" like features are observed in the temperature range 105~K <$ T $< 266~K along both directions (much stronger in the ab plane).This observation indicates the complex spin texture with the possibility of canted or ferrimagnetic ordering in F4GT, which need further investigation \cite{May2019}. With the further decrease in temperature, another transition is observed at around 100 K, which is recently reported as spin reorientation transition(SRT)\cite{Seo2020,Mondal2021}.\\
(4) In the proximity of the SRT, a clear hysteresis loop opens up along the $c$ direction in $M-T$ data of FCC and FCH magnetization measurements. The hysteresis loop's lower and upper closing temperatures are 85 K and 115 K  with a temperature span of $\Delta T\simeq$ 30 K. The peak/dip positions are centered at slightly different temperatures for FCC and FCH magnetization data, respectively. 
The prominent thermal hysteresis around  $T_\text{SRT}$ signifies the first-order nature of transition\cite{Moldover1971}. However, no thermal hysteresis is visible in the FCC and FCH magnetization data for $H\parallel ab$ direction. This kind of hysteresis can appear between two stable/meta-stable magnetic phases with slight energy differences and can be tuned or eliminated by applying a small external field. In this case,  1$<$$H$$\leq$10 kOe may be sufficient to eliminate the difference. 
The magnetic anisotropy energy at the $T_\text{SRT}$ determines the switching of magnetization, which results in the lowering of the magnetic moment along $ab$ plan ($H \parallel ab$). 
At high temperature region  ($T_\text{SRT}$ < $T$ <$T_\text{C}$), the magnetic easy axis lies in the $ab$-plane.  At temperatures, $T<T_\text{SRT}$, magnetic easy axis switches along the $c$ axis. Several unique features are reported to be accompanied by the SRT, like structural phase transition and magnetic anisotropy energy change due to lattice relaxation \cite{Sharma2006}. The $\frac{dM}{dT}|_{FCC}$ vs $T$  plot reveal that applied magnetic field has a significant influence  on  $T_\text{SRT}$ and it decreases from 100 K to 94 K with increasing applied magnetic field from 0.1 kOe to 1.0 kOe. The complex nature of the SRT in F4GT needs further study by using a neutron scattering to find the origin and actual magnetic interaction happening near $T_\text{SRT}$.\\
(5) The difference between ZFC and FCC start appearing below \textit{T}$<$$T_\text{C}$ and becomes prominent at low temperature below $T_\text{SRT}$. The difference between FCC and ZFC is quite large along the $c$ axis compared to along $ab$ plane due to the anisotropic response of the magnetic moment.
			
Figure. \textcolor{blue}{\ref{fig:MT}(c-d)} shows isothermal magnetization $M(H)$ at various temperature $(T)$ measured up to 50 kOe magnetic field along $H\parallel ab$ and $H\parallel c$ directions. Below $T_\text{C}$, the $M(H)$ increases rapidly at low fields and saturate at higher magnetic field as expected for ferromagnetic ordering. The saturation magnetic moment at $T$=16~K along the $ab$ plane is about 2.12 $\mu_B$/Fe. The low coercive field ($H_\text{C}$) indicates soft ferromagnetism similar to the bulk Fe$_3$GeTe$_2$(F3GT) \cite{Liu2018}.  The Rhodes-Wohlfarth ratio (RWR) is useful to identify the difference between an itinerant or a localized spin system. The RWR is defined as $P_c/P_s$, where $P_c$ is obtained from the effective moment $P_c(P_c$ + 2) = P$^2_{eff}$ and $P_s$ is the spontaneous magnetization in the ground state\cite{Wohlfarth1978,Moriya1979}. It is observed that for a localized system RWR = 1, and an itinerant system it is $\textgreater$ 1 \cite{Liu2017}. In this compound the obtained value is 2.35 which is slightly less then the value obtained in Fe$_3$GeTe$_2$ (RWR = 3.8)\cite{Mondal2021}. Therefore the concentration of Fe plays a significant role in  Fe$_l$GeTe$_2$ (with $l$ = 3, 4 and 5) for their itinerant characters\cite{Liu2017,Chen2013}.  
	\begin{figure}
	\centering
	\includegraphics[width=1\linewidth]{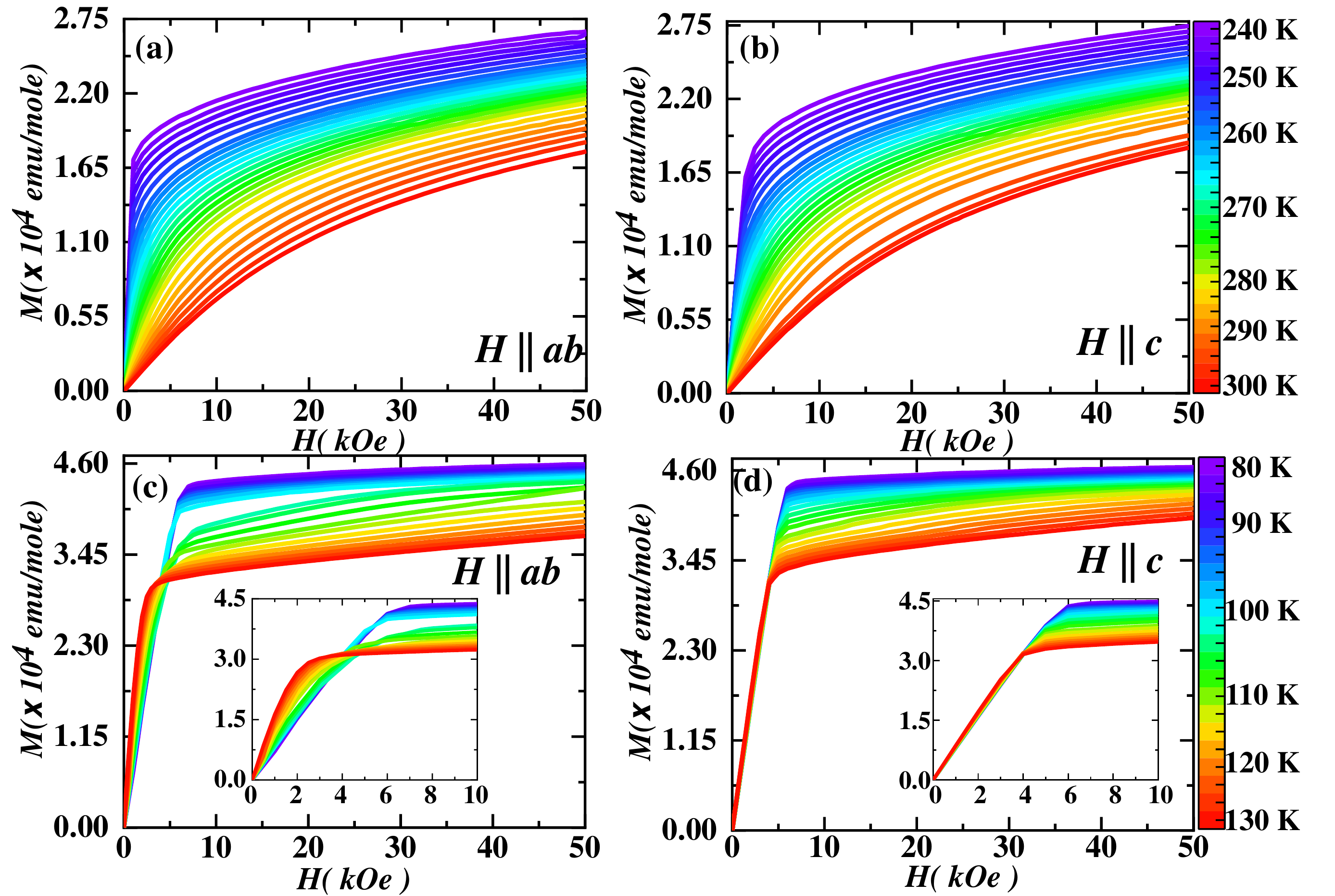}   	
	\caption{\textbf{Isothermal magnetization measurements.} The $M-H$ curves measured at around ferromagnetic transition temperature,  $T_\text{C}$ (\textbf{a-b)} and and spin reorientation transition temperature, $T_\text{SRT}$ \textbf{(c-d )} in directions of $H\parallel ab$ 
		and $H\parallel c$, respectively.}
	\label{fig:MH}	
\end{figure}

\subsection{Magnetocrystalline anisotropy}
\begin{figure}
		\centering
		\includegraphics[width=0.75\linewidth]{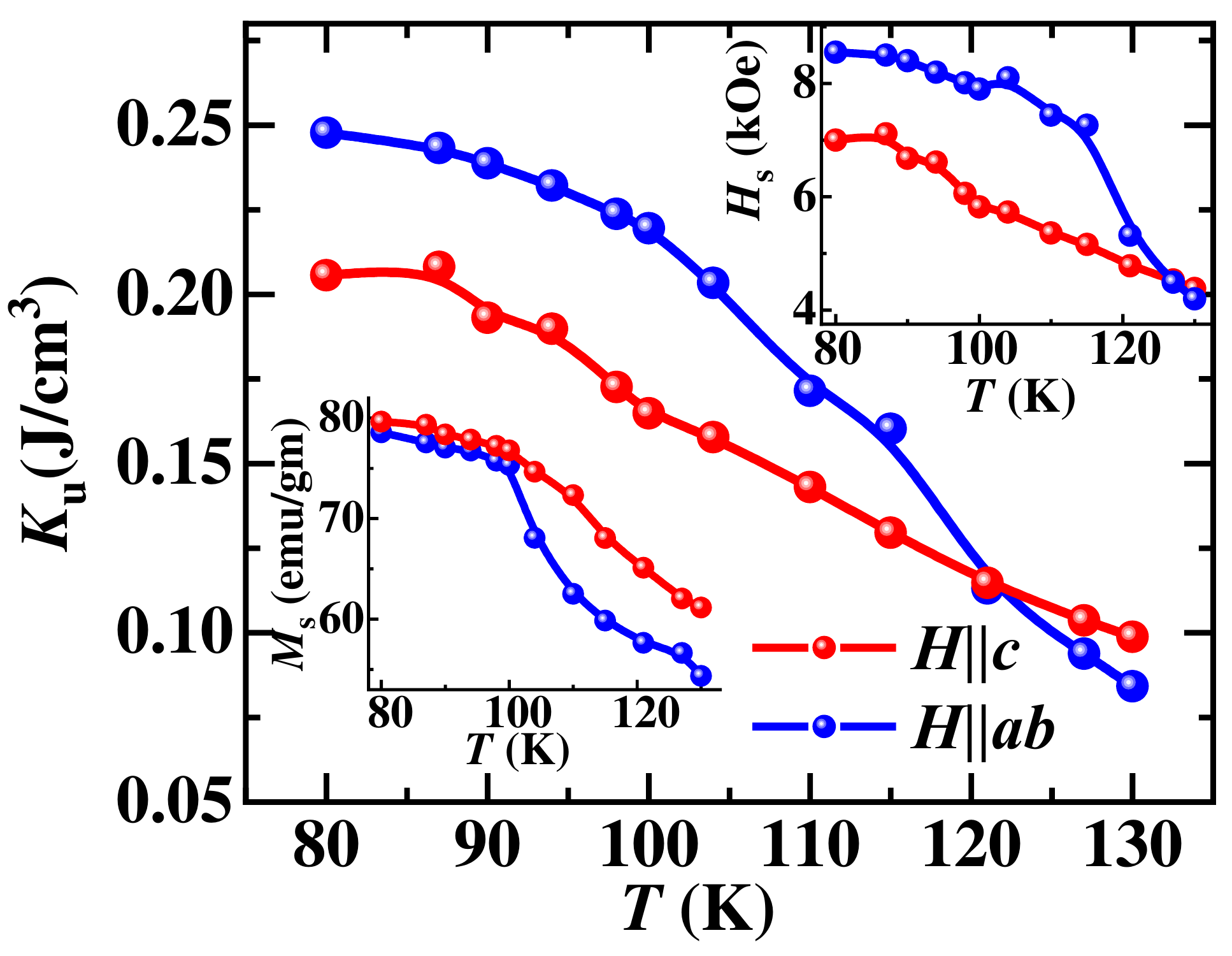} 		
		\caption{\textbf{Magnetic anisotropy at around SRT.} The estimated magnetic anisotropy constant, $K_\text{u}$, as function of temperature estimated from saturation field $H_\text{s}$, and the saturation
			magnetization$M_\text{s}$ (inset) around  $T_\text{SRT}$. }
		\label{fig:aniso}	
	\end{figure}

	~Magnetocrystalline anisotropy is an important property of ferromagnetic order in 2D materials, since the ferromagnetic order in 2D is vulnerable to the spin rotational fluctuations \cite{Mermin1966,Jin2018,Zhong2017}.~Uniaxial magnetocrystalline anisotropy constant ($K_\text{u}$), can be estimated from Stoner-Wolfarth model using relation, $K_\text{u}$ = $\frac{H_\text{s}M_\text{s}}{2}$. Here $H_\text{s}$ and $M_\text{s}$ are saturation field and saturation magnetization respectively. Figure. \ref{fig:MH} shows the virgin $M(H)$ curves at around $T_\text{C}$ and $T_\text{SRT}$ measured in both parallel to \textit{ab} and \textit{c} directions. Linear fit of $M(H)$ at higher field provides the value of $M_\text{s}$, whereas intersection point between linear fit to the low magnetic field and the high magnetic field data gives the $H_\text{s}$ value. The value of $K_u$  $\sim$  0.25 J/$cm^3$ which is much smaller (4 times ) than that for F3GT ($K_u$=1.03 J/$cm^3$). Previous reports suggest that in case of F4GT, $K_\text{u}$ originates from the competition between the magnetocrystalline anisotropy and the shape anisotropy \cite{Seo2020}. The significant change in $K_u$ indicates that the inter-layer exchange interaction is influenced by increasing  Fe concentration. The temperature-dependent $K_\text{u}$, $H_\text{s}$ and $M_\text{s}$ are presented in Figure. \textcolor{blue} {\ref{fig:aniso}}. These are strongly temperature dependent and the magnitude of $K_\text{u}$ at 80 K is 0.25 and 0.2 J/cm$^3$ along $H\parallel ab$ and  $H\parallel c$ respectively.
	The temperature dependence of the $K_\text{u}$ and $H_\text{s}$ decreases monotonically from the highest value at 80 K. Here, we clearly observe a crossover in  $K_\text{u}$ and $H_\text{s}$ around $T_\text{SRT}$, which signifies that the spins reorient from the direction of easy axis ($H\parallel c$) to the easy plane ($H\parallel ab$).
	
\subsection{Magnetocaloric effect (MCE)}
		
The MCE has been estimated from isothermal magnetization measured at temperatures around $T_\text{C}$ and $T_\text{SRT}$ (see Figure.\textcolor{blue} {\ref{fig:MH}(a,b)} and Figure.\textcolor{blue} {\ref{fig:MH}(c,d)}). The change in magnetic entropy ($\Delta$S$_M$) caused by a variation in the applied magnetic field ($H$) has been estimated using the following Maxwell relation \cite{Pecharsky1999},	

\begin{figure}
		\centering
		\includegraphics[width=1\linewidth]{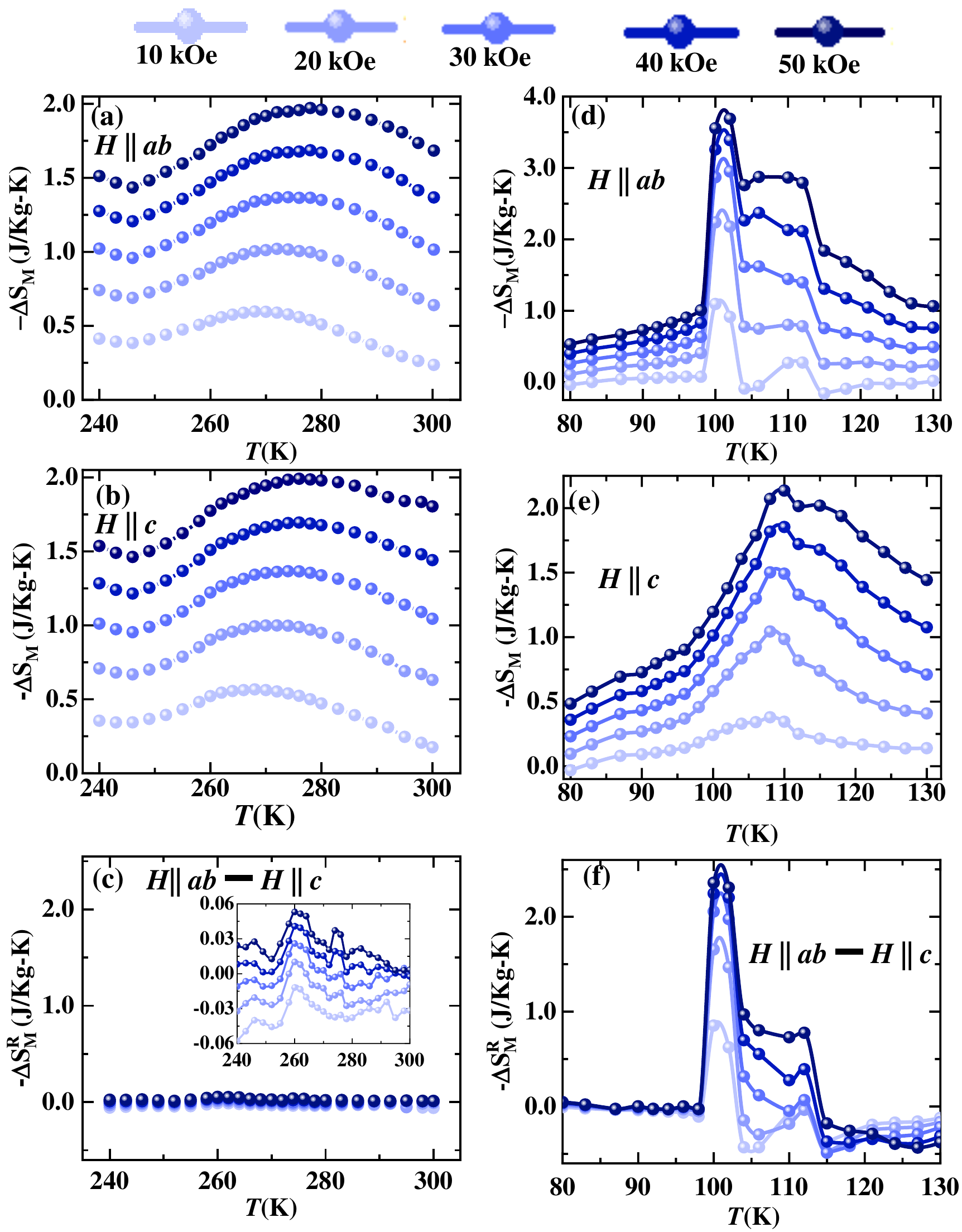}   	
		\caption{\textbf{Estimation of magneto caloric effect (MCE).}  Temperature-dependent isothermal magnetic entropy change -$\Delta$S$_M$ obtained from magnetization measurements at different
			magnetic fields in (a-b) at around $T_\text{C}$,  and (d-e) at around $T_\text{SRT}$  for $H\parallel ab$ 
			and $H\parallel c$, respectively. Temperature dependence of magnetic entropy change around $T_\text{C}$ \textbf{(c)} and $T_\text{SRT}$ \textbf{(f)} 
			obtained by subtracting from the $ab$ plane to the $c$ axis at various magnetic fields.}
		\label{fig:MCE}	
	\end{figure}
	
\begin{center}
	\begin{equation}
		\begin{split}
			\Delta S_M(T,H)=\int_{0}^{H} \frac{\delta M(T,H)}{\delta T} \,dH \
			\label{eq:mce}
		\end{split}
	\end{equation}
\end{center}
Figure~\textcolor{blue}{\ref{fig:MCE}(a,b and d,e)} show the estimated $\Delta S_M$ using the equation~\ref{eq:mce} in proximity to the ferromagnetic and spin reorientation transition. The isothermal magnetic entropy change curves (-$\Delta S_M$ vs $T$ plots) show   a broad maxima at around FM transition temperature $T_C$ along the both $c$ axis and $ab$ plane. Whereas -$\Delta S_M$ sharply rises to maxima along the $ab$ plane and shows  slightly broader asymmetric peak along the $c$ axis at SRT temperature, $T_\text{SRT}$. The maxima of $\Delta S_M$ monotonically increases with increasing applied magnetic field, $H$. For the magnetic field change of $\Delta H$= 50 kOe, the maximum entropy changes at $T_\text{C}$ are 1.95 J.Kg$^{-1}$K$^{-1}$ and 1.99 J.Kg$^{-1}$K$^{-1}$ along $H\parallel ab$ and $H\parallel c$ direction, respectively. The maximum entropy change at $T_\text{SRT}$ are 3.9 and 2.4 J.Kg$^{-1}$K$^{-1}$ along respective directions and much larger than the corresponding values for FM transition at $T_\text{C}$ (see Figure. \textcolor{blue}{\ref{fig:MCE}(d,e)}). The amount of magnetic asymmetry is presented through magnetic rotating entropy change ($\Delta S_M^R$) and can be expressed as,
\begin{center}
	\begin{equation}
		\begin{split}
			\Delta S_M^R(T,H)=\Delta S_M(T,H_c)-\Delta S_M(T,H_{ab})
		\end{split}
		\label{eq:dels}
	\end{equation}
\end{center}
 Unlike Fe$_3$GeTe$_2$ (F3GT), no such prominent asymmetry or change has been observed near the FM transition temperature $\sim$ 266 K in $\Delta$S$_M^R$ (see Figure. \textcolor{blue}{\ref{fig:MCE}(c)}). While in the vicinity of SRT, $\Delta$S$_M^R$ has been found to be strongly asymmetric and temperature-dependent (see Figure. \textcolor{blue}{\ref{fig:MCE}(f)}). In a second order phase transition, the existence of short-range order and spin fluctuations just above $T_\text{C}$ reduces the  |$(\frac{dM}{dT})_H$| and consequently the MCE. However, in a first order phase transition the value of |$(\frac{dM}{dT})_H$| should ideally be infinitive as it changes discontinuously at a constant temperature. In a real system like this, the sharp nature of MCE at $T_\text{SRT}$ compared to the broad nature at $T_\text{C}$ is consistent to the order of phase transition at both the temperatures. The sharp enhancement in MCE at $T_\text{SRT}$ between two crystalline directions firmly indicates towards the strong effect of magnetocrystalline anisotropy and first order nature of SRT \cite{Giguere1999,Casanova2003}. The magnetocaloric materials with first-order magnetic phase transitions are desirable in magnetic cooling technology, as these transitions are usually accompanied by a large change of magnetic entropy \cite{Skokov2012}. The magnetic refrigerants cooling efficiency is expressed in relative cooling power (RCP), which corresponds to the amount of heat moved between the cold and hot sinks in the ideal refrigeration cycle presented by\cite{GschneidnerJr.1999},
\begin{center}
	\begin{equation}
		\begin{split}
			RCP=|\Delta S_M^{max}|\times\delta T_{FWHM}
		\end{split}
	\label{eq:rcp}
	\end{equation}
\end{center}
Here, $\Delta$S$_M^{max}$ the maximum of entropy change and $\delta T_{FWHM}$ is value of $\Delta$S$_M$ at full width at half maximum. At around $T_\text{C}$, the estimated maximum RCPs are 81 and 82 in the $H\parallel c$ and $H\parallel$b direction at  5 T and near  $T_\text{SRT}$ the obtained values are 56 and 57. The values of RCP is in the moderate regime with reference to the known magnetic cooling technology \cite{Ding2013,Provenzano2004,Gorsse2008}. 

	\subsection{Critical scaling analysis of MCE}	

\begin{figure}
	\centering
	\includegraphics[width=.85\linewidth]{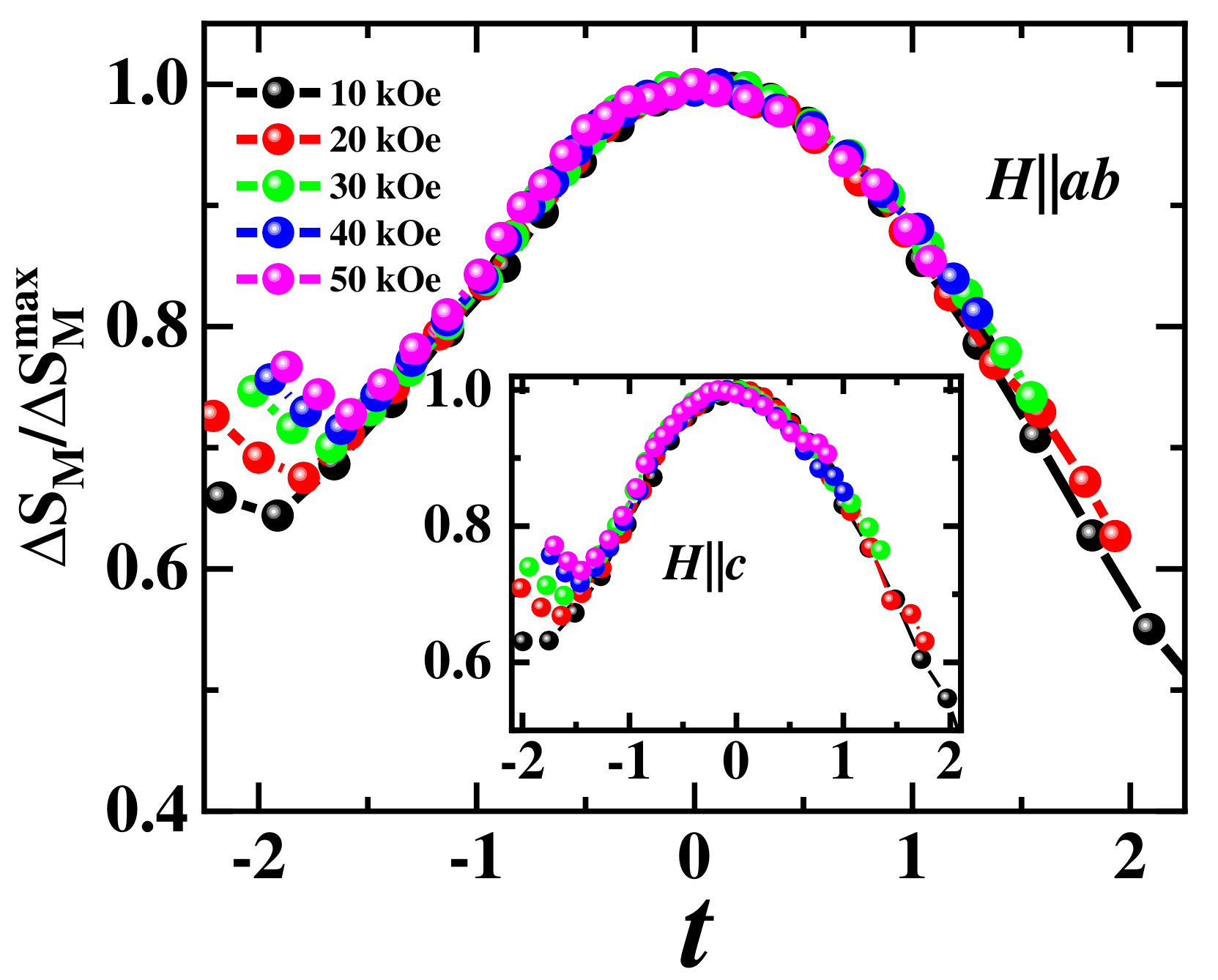}   	
	\caption{\textbf{Scaling analysis of MCE in proximity to FM transition at $T_{\text{C}}$.} Normalized $\Delta$ S$_M$ vs $t$ along $H\parallel ab$ plane. Inset shows the similar for $H\parallel c$ axis.}
	\label{fig:MCE_Scale}	
\end{figure}

The critical scaling analysis of MCE can unambiguously reveal the order of phase transitions which is very important for understanding the fundamental nature of magnetic phase transitions in low dimensional systems \cite{Law2018}. The -$\Delta S_M(T,H)$  curves can be rescaled onto a single plot regardless of the applied magnetic field ($H$) temperature ($T$) \cite{Franco2010} in proximity to the second-order phase transition. In this model, the normalized entropy change{$\Delta$ S$_M$}/{$\Delta $S$_M^{max}$}  is plotted against rescaled temperature ($t$) given as,

	\begin{center}
		\begin{equation}
			\begin{split}
		t_-=\frac{(T_{peak}-T)}{(T_{r1}-T_{peak})},  T<T_{peak}\\
		t_+=\frac{(T-T_{peak})}{(T_{r2}-T_{peak})},  T>T_{peak}
			\end{split}
		\label{eq:normce}
		\end{equation}
	\end{center}

Here $T_{peak}$ is the peak temperature where $\Delta$S$_M(T,H)$ shows maximum value. $T_{r1}$ and $T_{r2}$ are the reference temperature at which $\Delta$S$_M(T_{r1})$= $\Delta$S$_M(T_{r2})$=$\frac{1}{2}$$\Delta$S$_M^{max}(T,H)$ and $T_{r1}<T_\text{C}<T_{r2}$. Figure.\textcolor{blue}{\ref{fig:MCE_Scale}} shows ($\Delta$S$_M$/$\Delta$S$_M^{max}$)  vs $t$ at different applied external magnetic fields in $H\parallel ab$ and $H\parallel c$ direction. The plots of rescaled entropy changes  merge into one curve irrespective of  $H$ and $T$, which confirms that the second order nature of FM phase transition\cite{Liu2018b}. However, same scaling analysis in the vicinity of SRT does not work (see the Figure \textcolor{blue}{\ref{fig:MCE_Scale_t_100}}), which implies that SRT is not a second order type phase transition.

The magnetic entropy change ($\Delta$S$_M$) follow the power law expression in the proximity of phase transition as \cite{Oesterreicher1984a,Franco2010a,Law2018},

\begin{center}
	\begin{equation}
		\begin{split}
		|\Delta S_M|\varpropto H^n
		\end{split}
	\label{eq:powerLaw}
	\end{equation}
\end{center}

	\begin{figure}
	\centering
	\includegraphics[width=1\linewidth]{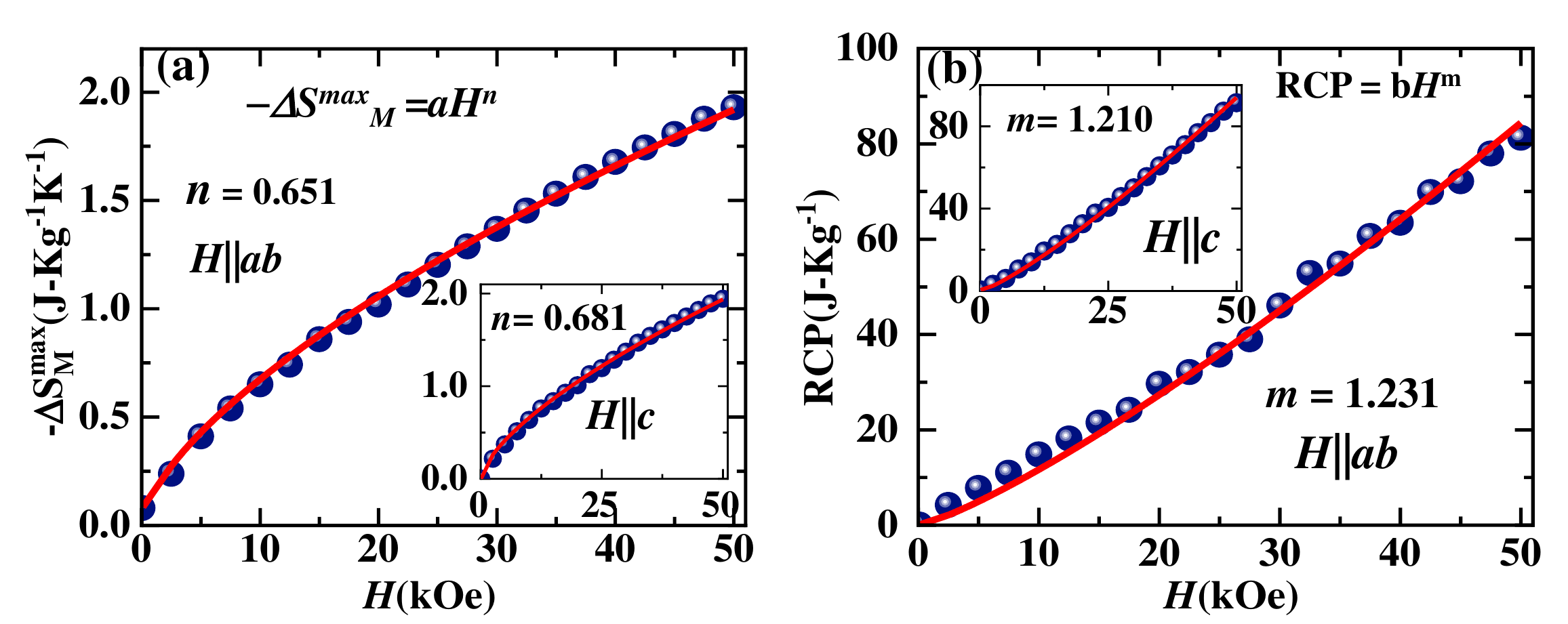}   	
	\caption{\textbf{Ferromagnetic transition at around $T_{\text{C}}$: Critical exponents. (a)}Magnetic field dependence of the maximum magnetic entropy change -$\Delta$S$^\text{max}_\text{M}$ along $ab$ plane (inset shows along the $c$-axis) and \textbf{(c)} The relative cooling power (RCP) with the power law fitting in red solid lines along $ab$ plane (inset shows along the $c$-axis).}
	\label{fig:MCE_Scale2}	
\end{figure}

Here, the exponent $n$ depends on temperature $(T)$, magnetic field $(H)$  and magnetic state of the compounds\cite{Li2008}. For estimation of $n$, the above equation~\ref{eq:powerLaw} is expressed as,

\begin{center}
	\begin{equation}
		\begin{split}
		n=\frac{\delta ln|\Delta S_M|}{\delta ln(H)}
		\end{split}
		\label{eq:powerconst}
	\end{equation}
\end{center}

The exponent \textit{n} in the equation~\ref{eq:powerconst}, is a crucial parameter for the identification of order phase transition (i.e. whether it belongs to first or second order phase transition )\cite{Bean1962,Law2018}.

\begin{figure}
	\centering
	\includegraphics[width=1\linewidth]{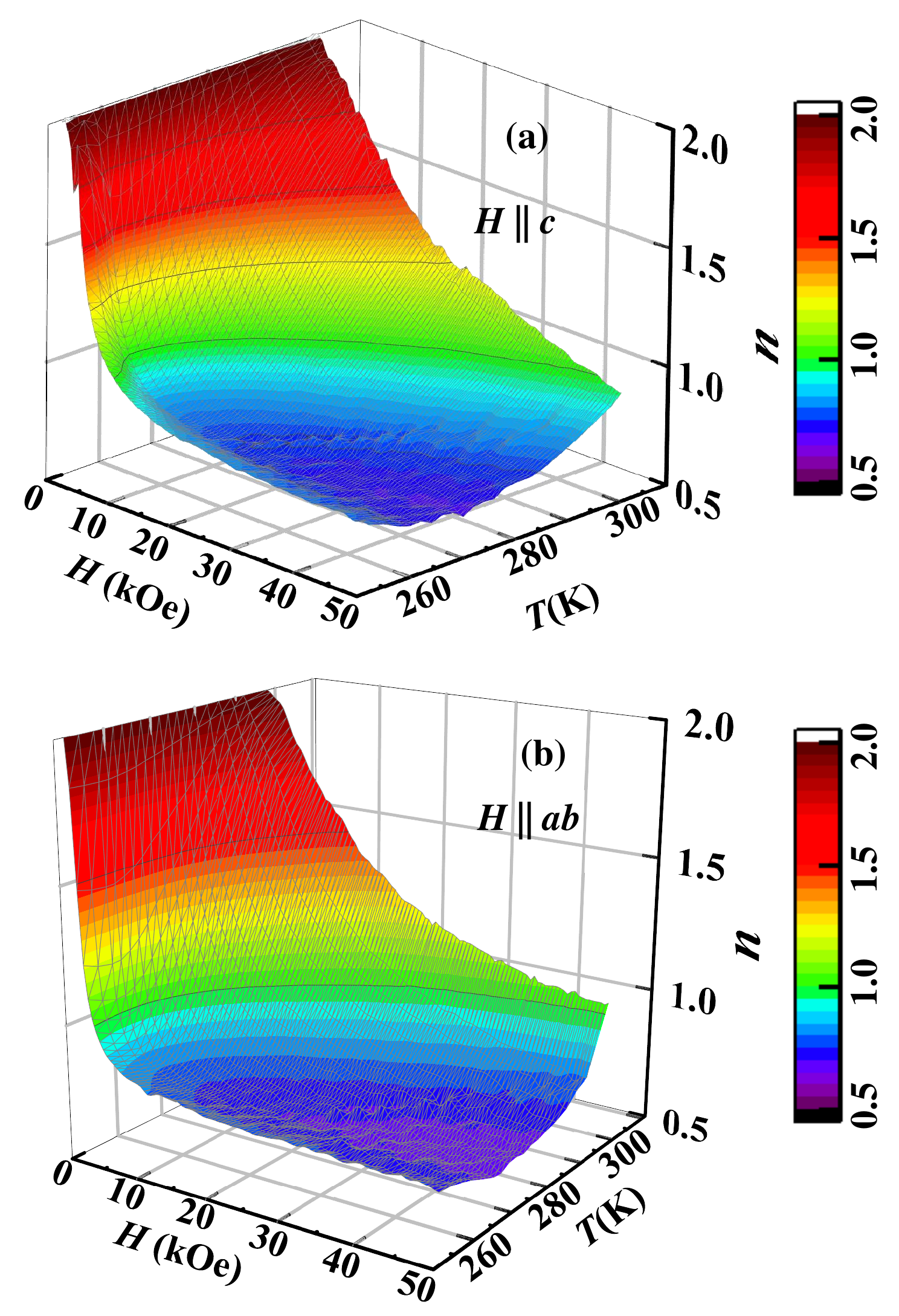}   	
	\caption{\textbf{The exponent $n$ in proximity to the FM and SRT transition:(a-b)} Temperature dependence of critical exponent, $n$ estimated using equation~\ref{eq:powerconst} in proximity to  ferromagnetic transition at around $T_{\text{C}}$ \protect \cite{Law2018}.}
	\label{fig:MCE_Scale_n270}	
\end{figure}

In case of second order phase transition, the $\Delta S_{M}^{max}\varpropto H^n$ with the exponent $n=2/3$ at the transition temperature predicted from the mean field model \cite{Oesterreicher1984a}. Interestingly, for the FM transition obtained values of $n$ are 0.65 and 0.68  for applied field along $ab$ and $c$ directions, which are very close to the predicted mean field  value, $2/3$ \cite{Oesterreicher1984a} (see the Figure~\textcolor{blue}{\ref{fig:MCE_Scale2}(a)}). Figure.\textcolor{blue}{\ref{fig:MCE_Scale_n270}} shows the variation of $n$ estimated using the expression~\ref{eq:powerconst} in proximity to FM and SRT transition.~The 3D surface plots in Figure \textcolor{blue}{\ref{fig:MCE_Scale_n270}(a and b)} clearly shows that  value of $n$ exhibit minima  at $T_\text{C}$ and approaches 1 well below and 2 well above the $T_\text{C}$ along both directions as expected for second order phase transition \cite{Law2018}. 
 \begin{figure}
 	\centering
 	\includegraphics[width=1\columnwidth]{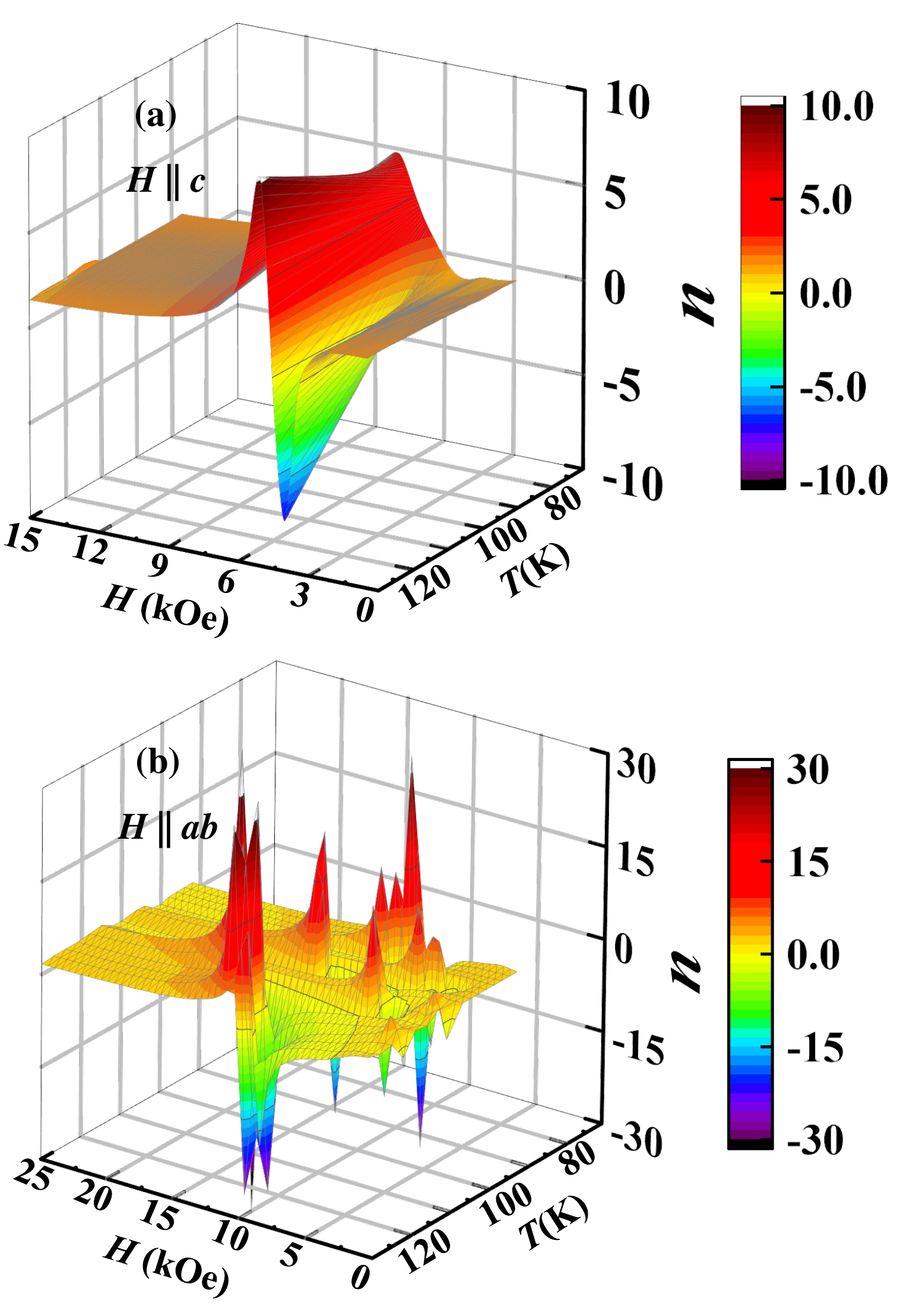}   	
 	\caption{\textbf{The exponent $n$ in proximity to the SRT: (a-b)} Temperature and field dependence of critical exponent, $n$ estimated using equation~\ref{eq:powerconst} in proximity to SRT. The figure shows that $|n|$ become larger than 2 at around $T_{\text{SRT}}$ that confirm the first oder nature of the SRT \protect \cite{Law2018}}. 
 	\label{fig:MCE_Scale_n100}	
 \end{figure}
For a first order thermo-magnetic phase transition, value of |\textit{n}| is expected to be greater than 2 at around the transition temperature\cite{Bean1962,Law2018}. In this study, the estimated value of |$n$| at around $T_\text{SRT}$ is much larger than 2 which suggest that SRT is a first order phase transition consistent with other thermodynamic measurements such as magnetization (see Figure~\textcolor{blue}{\ref{fig:MCE_Scale_n100}}). In addition, the maximum entropy change at around $T_\text{SRT}$ does not follow the power law behavior ($S_{M}^{max}\varpropto H^n$) as expected for the second order phase transition \cite{Law2018}.The breakdown of power law has bean demonstrated in Figure~\textcolor{blue}{\ref{fig:MCE_Scale_t_100}}. 
 
The exponent $n$ is related to the critical exponents $\beta$, $\gamma$ and $\delta$ in proximity to second order phase transition through the following relation~\cite{Oesterreicher1984a,Franco2010}
 	\begin{center}
 	\begin{equation}
 		\begin{split}
 		n|_{T=T_c}=1+\frac{(\beta-1)}{(\beta+\gamma)}=1+\frac{1}{\delta}(1-\frac{1}{\beta})
 		\end{split}
 	\label{eq:exponent}
 	\end{equation}
 \end{center}
Similarly, RCP related to the applied magnetic field through the equation,
	\begin{center}
	\begin{equation}
		\begin{split}
			RCP=bH^m
		\end{split}
	\label{eq:rcp1}
	\end{equation}
\end{center}

\begin{figure}
 	\centering
 	\includegraphics[width=.85\linewidth]{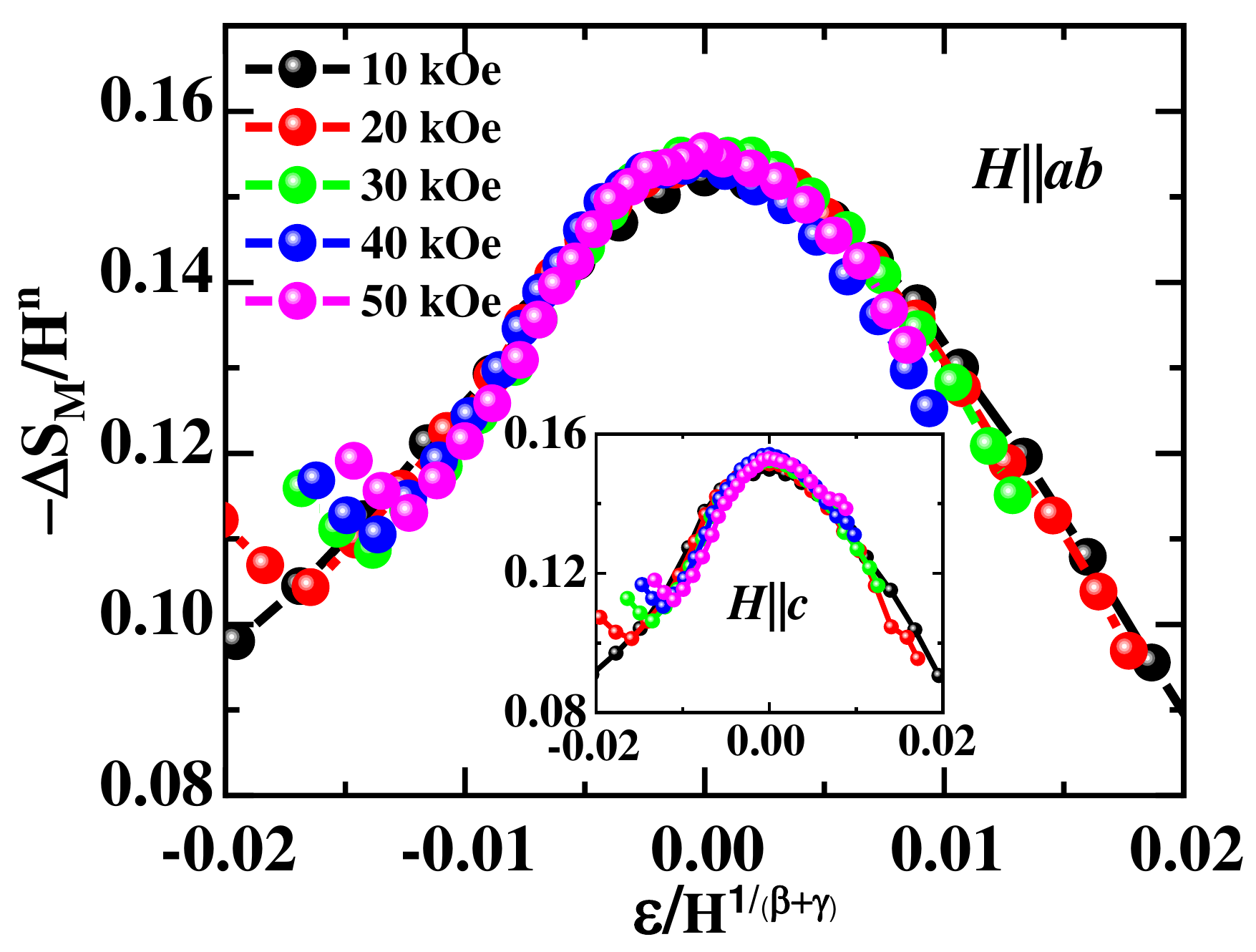}   	
 	\caption{\textbf{Scaling analysis of MCE in proximity to FM transition at $T_{\text{C}}$.} Scaling plot based on the critical exponents $\beta$ and $\gamma$. Inset shows the similar for $H\parallel c$ axis.}
 	\label{fig:MCE_Rescale}	
 \end{figure}

Here $b$ is a constant and $m$ is related to critical exponent $\delta$  by the simple formula, $m$=1+ $\frac{1}{\delta}$. The fitting of RCP values as shown in Figure. \textcolor{blue} {\ref{fig:MCE_Scale2}(c)} gives $m$=1.23 and 1.21 along $ab$ and $c$ directions respectively. Using the equation \ref{eq:exponent} the values of other two critical exponents, $\beta$ and $\gamma$ are estimated and presented in table \ref{table:nonlin}. 
For second order phase transition, -$\Delta$S$_M$  can also be expressed  using the following scaling relation  and demonstrated  in Figure.\textcolor{blue} {\ref{fig:MCE_Rescale}} \cite{Su2013},
\begin{center}
	\begin{equation}
		\begin{split}
		\frac{-\Delta S_M}{a_M}=H^nf(\frac{\epsilon}{H^{\frac{1}{\Delta}}})
		\end{split}
	\label{eq:equation8}
	\end{equation}
\end{center}
Here, a$_M$={A$^{\delta+1}$B}/{$T_\text{C}$}, where A and B are the critical amplitudes which are estimated using following equations,
\begin{center}
	\begin{equation}
		\begin{split}
			M_s(T)=A(\epsilon)^\beta , H=BM^\delta
		\end{split}
	\label{eq:equation9}
	\end{equation}
\end{center}
 Here, $\Delta$=$\beta$+$\gamma$, and $f(x)$ is the scaling function. By proper choice of critical exponents, the resealing of reduced temperature and magnetic entropy change should  converge all experimental  -$\Delta$S$_M$ on a single curve.  
 
  \begin{table}
 	\caption{Critical exponents of F4GT compared with few experimental model mentioned in \cite{Mondal2021}. (MAP = modified Arrott plot, KF = Kouvel Fisher plot and CI = critical isotherm)}. Upper and lower data correspond to $ab$ and $c$ direction respectively. 
 	
 	\begin{tabular}{c c c c c c c c}
 		\hline\hline 
 		Techniques & Direction & $n$ & $m$ & $\beta$ & $\gamma$ & $\delta$  &  Reference \\  [0.4ex] 
 		\hline\\ 
 		-$\Delta S_M^{max}$ & H||ab & 0.65 &  & 0.40 & 1.302 & & This work  \\ 
 		& H||c & 0.68 &  &  0.396 & 1.49 & & \\ \\ 
 		RCP & H||ab &  & 1.23 &  &  & 4.23 & This work   \\      
 		&H||c &  &1.21  & &  & 4.76  &             \\ \\ 
 		MAP &H||ab & & & 0.37 & 1.21 & 4.24 & \cite{Mondal2021} \\
 		&H||c & & & 0.33 & 1.17 & 4.54 &         \\ \\ 
 		KFP &H||ab  &  &  & 0.376 & 1.20 & 4.22& \cite{Mondal2021}    \\ 
 		& H||c & & & 0.33 & 1.20 & 4.63 \\ \\
 		CI &H||ab  & & &  &  & 4.21 & \cite{Mondal2021} \\          
 		& H||c & & &  &  & 4.52\\ \\  [1ex]
 		\hline    
 	\end{tabular} 
 	\label{table:nonlin} 
 \end{table}

 \textcolor{blue}{Figure~\textcolor{blue} {\ref{fig:MCE_Rescale}}} demonstrate the above mentioned scaling analysis  along $ab$ and $c$ direction at around FM transition, by taking $\beta$  and $\gamma$ values from earlier calculation mentioned in Table \textcolor{blue} {\ref{table:nonlin}}. The obtained value of critical constants for FM transition are not belong to any known universal class, however constants do satisfy the scaling relations.  The above scaling analysis of MCE clearly confirm the second order nature of FM phase transition in F4GT\cite{Franco2010,Law2018}.  

However, the same scaling analysis of MCE in proximity  to SRT breaks down as demonstrated in appendix~\ref{app:m_SRT}. The temperature and field dependence of critical exponent, $n$ in proximity to SRT in \textcolor{blue}{Figure~\ref{fig:MCE_Scale_n100}} confirms the first order nature of SRT \cite{Law2018}.Therefore, scaling analysis of MCE has unambiguously established that the SRT is first order phase transition.

\section{CONCLUSION}
 	We have investigated the nature of magnetic ordering through anisotropic magnetic entropy change in quasi-two-dimensional (2D) van der Waals ferromagnetic material, F4GT. The single-crystals of layered ferromagnet F4GT were grown using the chemical vapor transport method. The detailed study of magnetization and magnetic entropy changes clearly reveals that F4GT undergoes two distinct magnetic transitions: (1) paramagnetic to ferromagnetic at $T_\text{C}$ $\sim$ 266 K and another (2) spin reorientation transition at around $T_\text{SRT}$ $\sim $ 100 K. The SRT has also been observed in specific heat measurements. The magnetic easy axis of F4GT lies in the $ab$-plane in temperature region $T_\text{SRT}$ < $T$ <$T_\text{C}$ but gets switched along the $c$ axis below $T_\text{SRT}$ leading to the spin reorientation transition. Detailed critical scaling analysis confirms the second-order nature of ferromagnetic transition, whereas the SRT is identified as a first-order phase transition. Nonetheless, F4GT with higher FM transition temperature and the tuning capability of magneto-crystalline anisotropy allows us to manipulate the spins from in-plane to out-of-plane and provide a new platform for exploring low dimensional magnetism in layered 2D van der Waals magnetic materials.

	\section{ACKNOWlEDGEMENT}
	This work was supported by the (i) 'Department of Science and Technology', Government of India (Grant No. SRG/2019/000674 and EMR/2016/005437), and (ii) Collaborative Research Scheme (CRS) project proposal(2021/CRS/59/58/760). S.Bera $\&$ A.Bera thanks CSIR Govt. of India for Research Fellowship with Grant No. 09/080(1110)/2019-EMR-I  $\&$ 09/080(1109)/2019-EMR-I, respectively. S.B. acknowledges the experimental facilities for sample growth, procured using financial
	support from DST-SERB grant nos. ECR/2017/002 037. Authors also would like to acknowledge Dr. Subhadeep Datta and Prof. Subham Majumdar for fruitful discussions.
	
\appendix
\renewcommand{\thefigure}{A\arabic{figure}}

\setcounter{figure}{0}

\section{Elemental composition analysis of F4GT: SEM-EDX}
\label{apsec:SEM-EDX}

\begin{figure}
	\centering
	\includegraphics[width=1\linewidth]{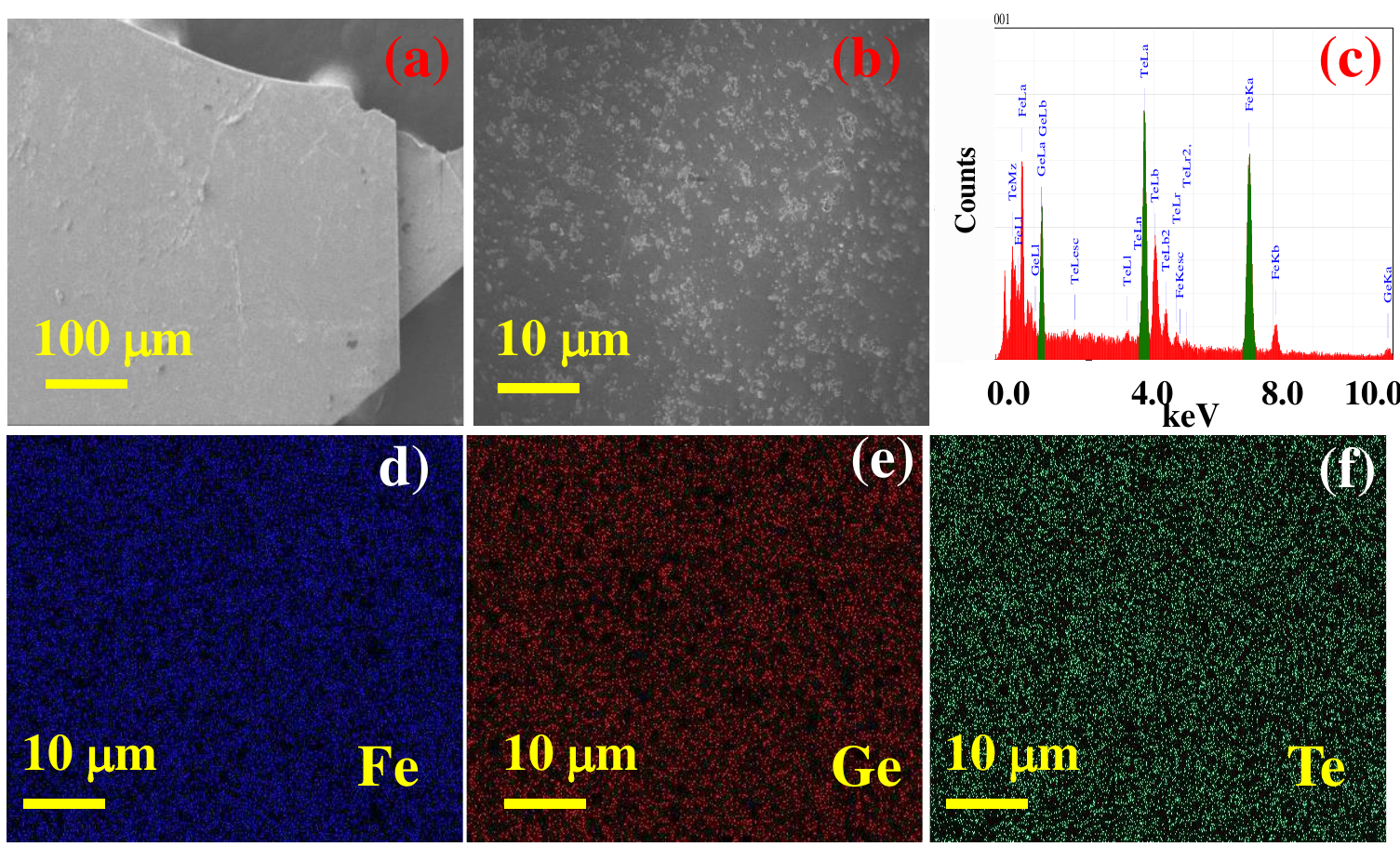}   	
	\caption{\textbf{Elemental analysis of F4GT.} (a-b) High resolution SEM image of this cryatal. (c) A typical energy-dispersive X-ray (EDX) spectrum of the this crystal and (d-f) EDX mapping with the elemental distribution of Fe, Ge and Te respectively.}
	\label{fig:SEM}	
\end{figure}

Figure.\textcolor{blue}{\ref{fig:SEM}(a-b)} show low-magnification FESEM image of the F4GT single crystal. The layered structure of the sample is clearly visible here. To examine the elemental composition and chemical homogeneity, energy-dispersive X-ray (EDX) spectroscopy on the crystal has been carried out(See Figure.\textcolor{blue}{\ref{fig:SEM}(c)}). The elemental distributions of Iron (Fe), Germanium (Ge) and Tellurium (Te) obtained from SEM-EDS, are shown in Figure.\textcolor{blue}{\ref{fig:SEM}(d-f)}, which confirms that the constituent elements are evenly distributed. The obtained atomic weight percentage of the Fe,Ge, and Te are 59.4, 13.5, and 27.3, confirming the presence of Fe:Ge:Te in a ratio of 4.18:0.95:1.92 without other impurities and matches the earlier report\cite{Mondal2021}.

\section{Magnetization of F4GT sample}
\label{app:Magnetization}		
\begin{figure}
		\centering
		\includegraphics[width=0.75\linewidth]{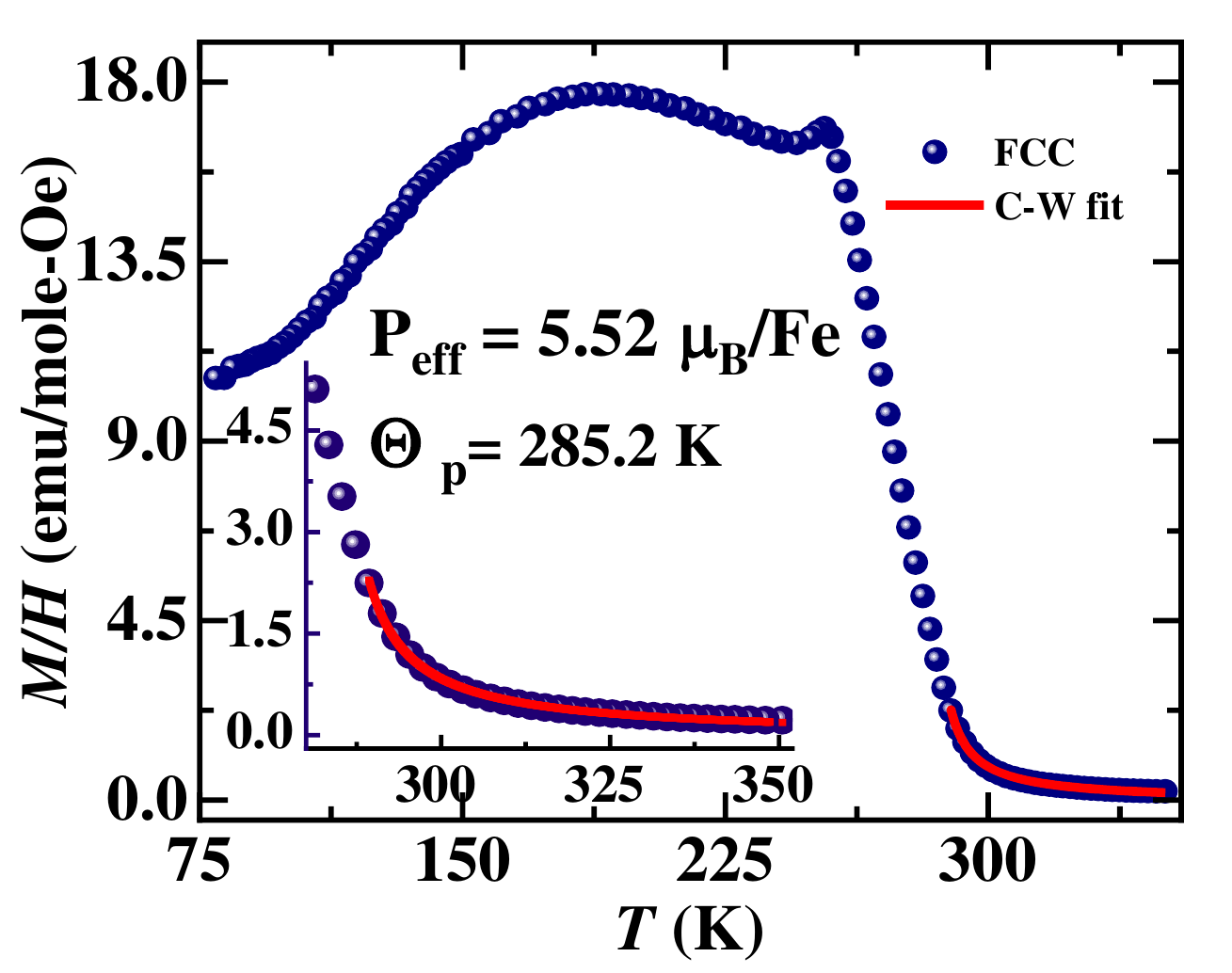}   	
		\caption{\textbf{The Curie-Weiss (CW) fit of MT data.} The $M/H$ vs $T$ of powder sample of F4GT and the corresponding modified Curie-Weiss (CW) fit.}
		\label{fig:MT_MH}	
	\end{figure}

 To estimate the effective magnetic moment of Fe atom, the $M/H$ vs $T$ of powder sample of F4GT has been analyzed using the  modified Curie-Weiss (CW) expression,
	\begin{center}
		\begin{equation}
			\begin{split}
				\frac{M}{H}=\chi_0+\frac{C}{(T-\Theta_p)}  \\
			\end{split}
			\label{eq:CW}
		\end{equation}
	\end{center}
Here, $C$ is the Curie constant and $\Theta_p$ is CW temperature. The temperature dependent magnetization data from 350K down to 80K has shown in Figure~\textcolor{blue}{\ref{fig:MT_MH}}.The fits yield effective magnetic moment, $P_\text{eff}$ = 5.52 $\mu_B$/Fe with Curie-Weiss temperature 285.2 K (see the Figure~\textcolor{blue}{\ref{fig:MT_MH}}). The +ve value of $\Theta_p$ confirms the FM characteristics of F4GT.

\section{Heat capacity measurement}
\label{app:heatcapacity}
\begin{figure}
	\centering
	\includegraphics[width=0.75\linewidth]{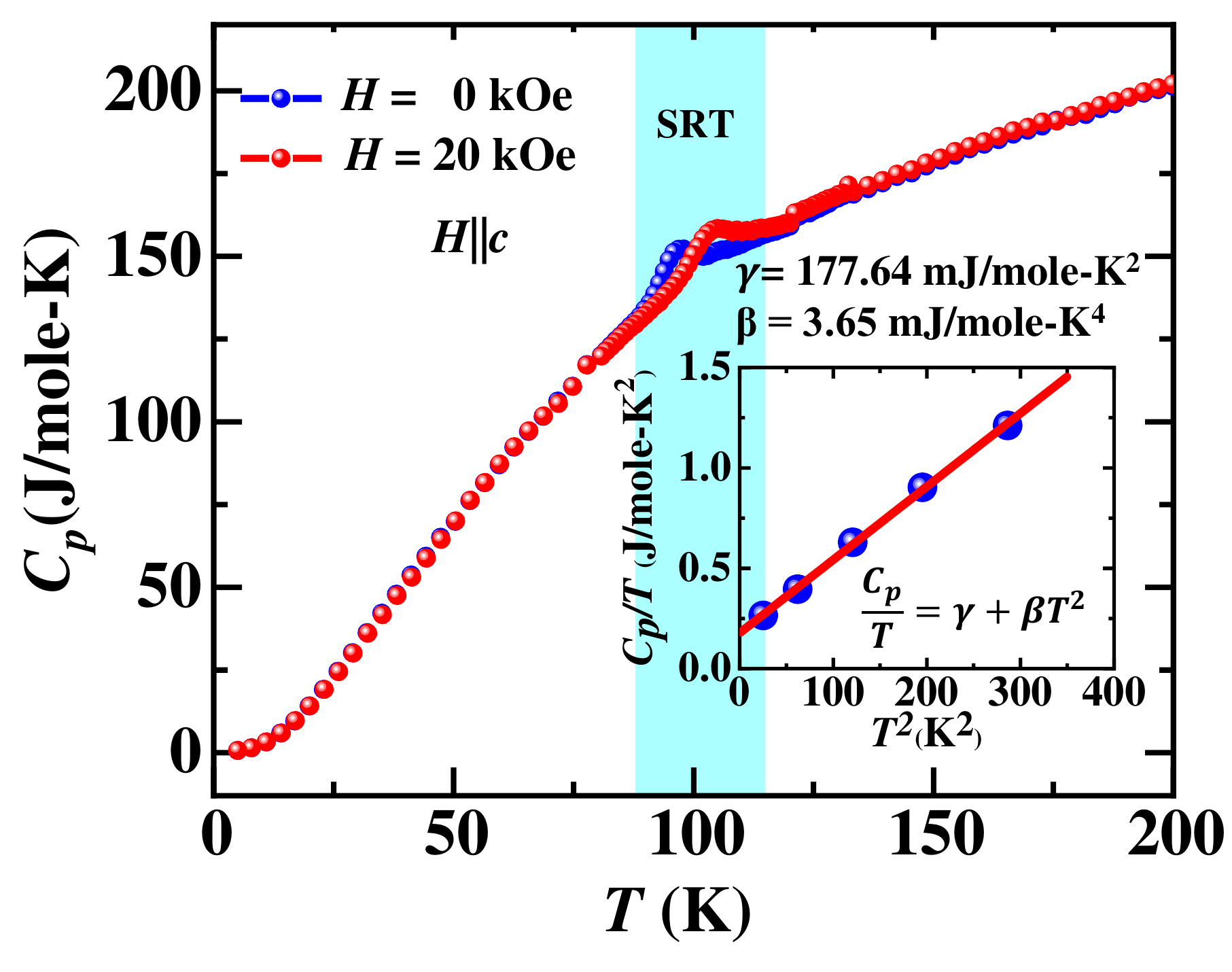}   	
	\caption{\textbf{Specific heat data.} Specific heat of Fe$_{4}$GeTe$_2$ under $H$ = 0 and 20 kOe magnetic fields over 5 - 200 K temperature range.}
	\label{fig:sp}	
\end{figure}

To get further insight of the spin-reorientation transition (in view of it's thermodynamic nature), we have measured specific heat ($C_\text{p}$) in the temperature range 5 - 200 K, without and with the application of external magnetic field $(H)$ (see Figure~\textcolor{blue}{\ref{fig:sp}}). The $C_\text{p}$ gradually decreases as $T$ decreases down to 10 K as expected from phonon contributions. Interestingly, it shows a prominent well defined anomaly/peak at around $T_\text{SRT}$ due to the magnetic ordering (spin reorientation transition)(see Figure~\textcolor{blue}{\ref{fig:sp}}). The anomaly/peak position is shifted with the applied magnetic field $H$, which reconfirms the magnetic origin of the anomaly/peak. The SRT is associated with the entropy change in this system and thus contributes to the specific heat, which appears as an anomaly/peak on top of the phonon contribution ($\propto T^3$) at around $T_\text{SRT}$.

The $C_\text{p}/T$ versus $T^2$ data show linear behavior between 5 and 16 K (Inset of Figure~\textcolor{blue}{\ref{fig:sp}}) implying a $\gamma$$T$+$\beta$$T$$^3$ type variation of total $C_\text{p}$ at low temperatures. For an ordinary metal, $\gamma$$T$ and $\beta$$T$$^3$ are, respectively, the electronic and lattice contributions to the heat capacity. From the linear fit to the data below 16 K, we obtain the Sommerfeld coefficient $\gamma$ = 177.64 mJ/mol-K$^2$
and $\beta$ = 3.65 mJ/mol-K$^4$. The large value of $\gamma$ suggests the present of significant electronic correlation in F4GT\cite{Zhu2016}. 
	
\section{-$\Delta$ S$_M$ vs H for fixed T}
\label{app:magnetEntropyChaneg}

\begin{figure}
	\centering
	\includegraphics[width=1\linewidth]{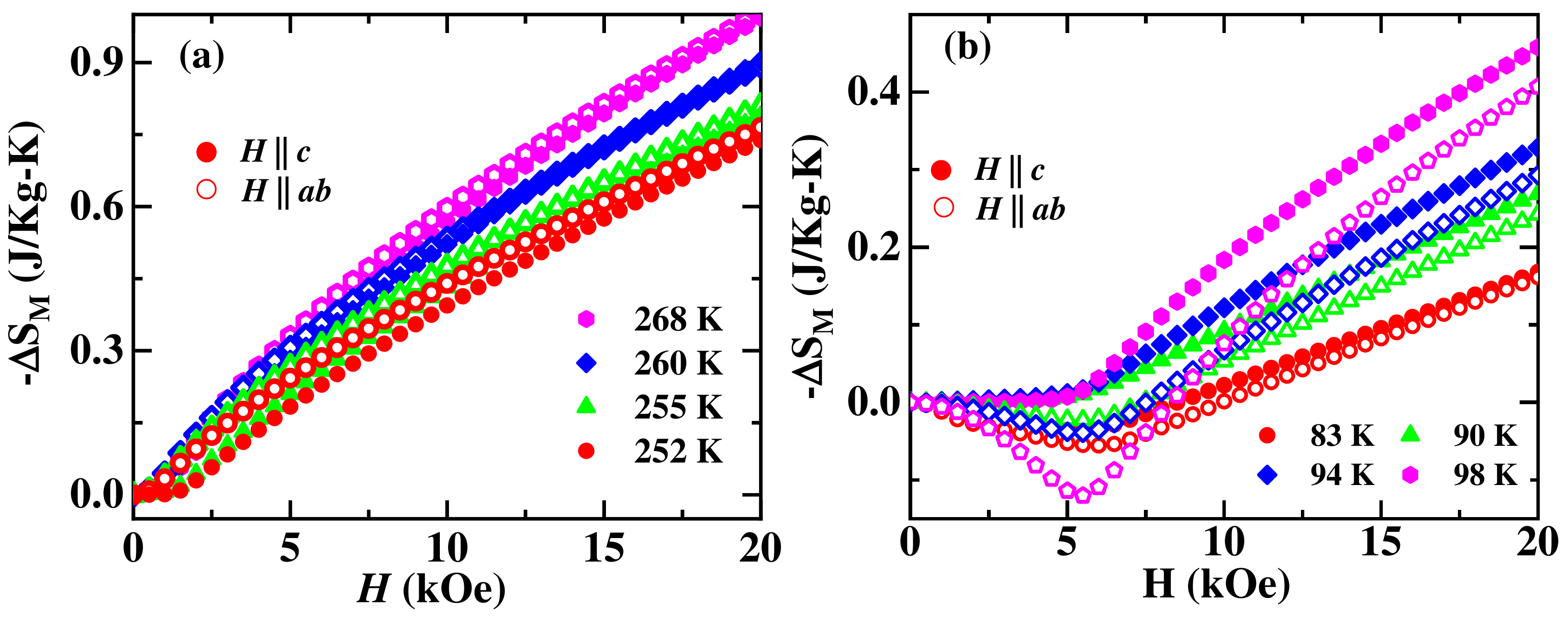}   	
	\caption{ \textbf{Estimation of magnetic entropy change with magnetic field.}  Field dependence of -$\Delta$S$_\text{M}$ at around(a) $T_\text{C}$ and (b)T$_{SR}$ along \textit{c}-axis(solid) and \textit{ab} plane (hollow) respectively.}
	\label{fig:iso}	
\end{figure}

	The field-dependent -$\Delta$ S$_M$ is shown in Figure.\textcolor{blue} {\ref{fig:iso}(a,b)}, in two transition temperature region. It is interesting to mention that around $T_\text{C}$ the values of -$\Delta$ S$_M$ along the both directions, i.e. ab plane and c axis are positive with very little difference, suggesting negligible amount of anisotropy. Where as around $T_\text{SRT}$, -$\Delta$ S$_M$  are negative at low fields along  the ab plane and all values are positive along the c axis, suggesting a considerable amount of  anisotropy.

\section{The scaling analysis of MCE in proximity to SRT}
\label{app:m_SRT}
\begin{figure}
	\centering
	\includegraphics[width=1\linewidth]{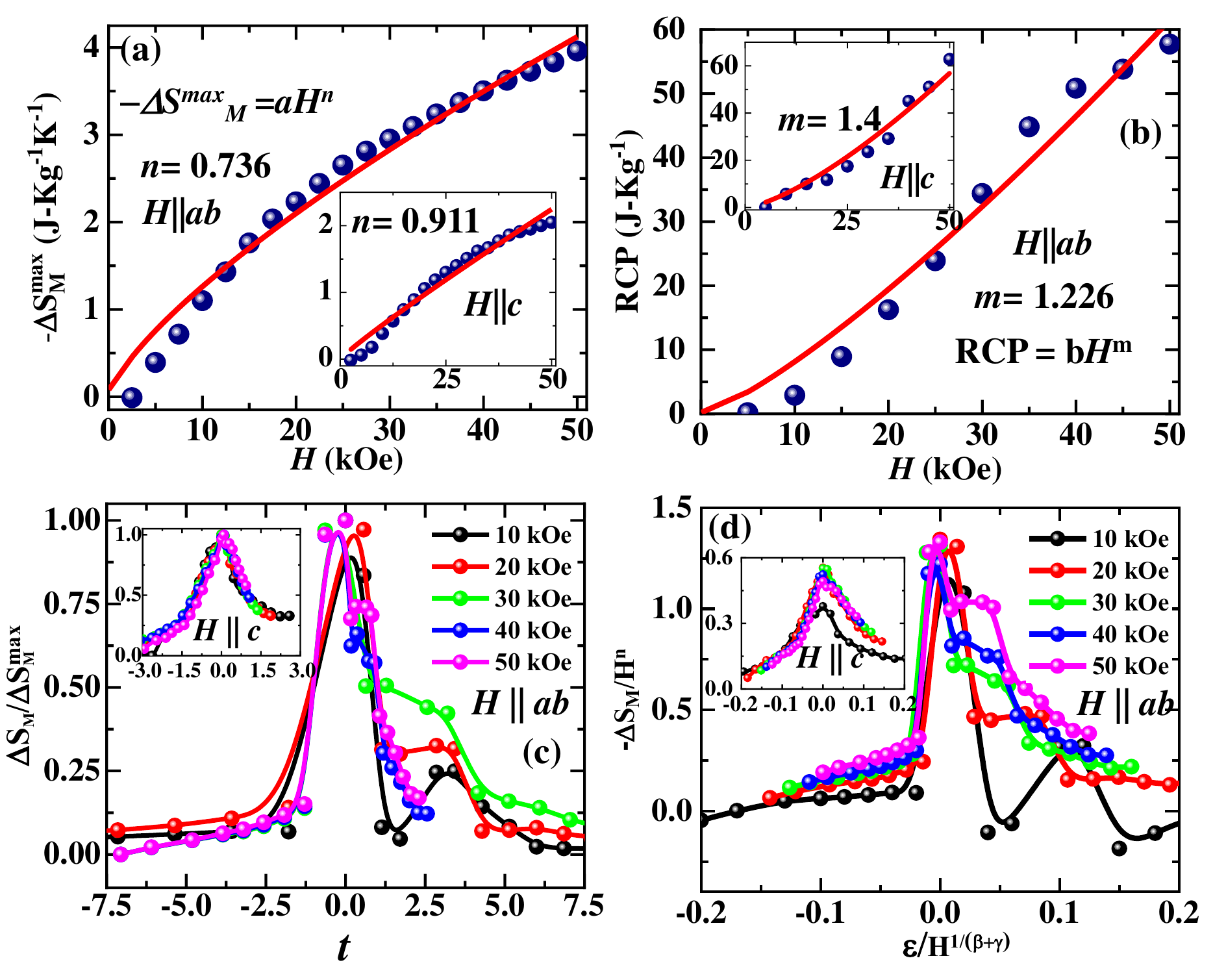}	
	\caption{{\textbf{Scaling analysis of MCE in proximity to spin orientation transition.} (a)} Magnetic field dependence of the maximum magnetic entropy change -$\Delta$S$^\text{max}_\text{M}$ along $ab$ plane (inset shows along the $c$-axis) and (b) The relative cooling power (RCP) with the power law fitting in red solid lines along $ab$ plane (inset shows along the $c$-axis) (c) Normalized $\Delta$ S$_M$ vs $t$ along $H\parallel ab$ plane. Inset shows the similar for $H\parallel c$ axis. (d) Scaling plot based on the critical exponents $\beta$ and $\gamma$.}
	\label{fig:MCE_Scale_t_100}	
\end{figure}

Figure~\textcolor{blue}{\ref{fig:MCE_Scale_t_100}} demonstrates the breakdown of scaling analysis of MCE at around spin reorientation transition. Here the obtained values of $n$ are 0.736 \& 0.911 and $m$ are 1.4 \& 1.226 along $ab$ and $c$-axis respectively. The rescaled $\Delta$S$_M (T, H)$ does not follow a universal curve confirming the deviation from second-order character. Based on the above estimated values of exponents $n$ and $m$, the rescaled $\Delta$S$_M (T, H)$ fails to converge into a single curve. These breakdown of the scaling analysis imply that SRT is not a second order transition\cite{Law2018}. 	

\bibliography{SBF4GT}	
\end{document}